\documentclass[amssymb,superscriptaddress,twocolumn]{revtex4}
\usepackage{graphicx}

\newcommand{\beq}[1]{\begin{eqnarray}\label{#1}}
\newcommand{\eeq}{\end{eqnarray}}

\begin{document}
%

%\today

\preprint{IMSc/2004/07/28}

%\title{Absence of particle horizon in semi-classical Loop Quantum Cosmology}
\title{
 Primordial Power Spectrum's Calculation\\
 and the Upper Bound\\
 on the Number of E-foldings of Inflation}

\author{Ding-fang Zeng and Yi-hong Gao}
\email{dfzeng@itp.ac.cn} \email{gaoyh@itp.ac.cn}
\affiliation{Institute of Theoretical Physics, Chinese Academy of
Science.}

\begin{abstract}
In numerical calculations of the primordial power spectrum of
perturbations produced during inflation, for very little
wavenumber $k$s, to implement the initial conditions required of
the perturbation mode functions consistently, we must let the
universe experience more e-foldings of inflation than that
required of it to solve the horizon, flatness and other
pre-inflation problems. However, if the number of e-foldings of
inflation has an upper bound, then for perturbations at scales
greater than some critical value, the initial conditions required
of the mode functions cannot be implemented physically. Because at
the inflation beginning point, these perturbation modes lie
outside the horizon. We propose that such perturbations do not
contribute to the Cosmological Microwave Background Anisotropy
(CMBA). Under this proposition, the exceptional lowness of the
observed little $l$ muli-poles of CMBA is reproduced numerically.
In Linde's $\phi^2$ model, the upper bound on the number of
e-foldings of inflation is determined to be 65 approximately.

\end{abstract}

%\pacs{04.60.Pp, 04.60.Kz, 98.80.Jk}

\maketitle

%%%%%%%%%%%%%%%%%%%%%%%%%%%%%%%%%%%%%%%%%%%%%%%%%%%%%%%%%%%%%%%%%%%%%%%
 \section{Introduction}
 Although calculations of the primordial power spectrum of
 perturbations produced during inflation have been discussed by
 several authors \cite{StewartLyth, Salopek, WenbingLin}, we still
 think that there are problems worthy of notifying in this
 paper. For example, in the slow roll approximations, to get
 the nearly scale free primordial power spectrum of perturbations,
 reference \cite{StewartLyth} and \cite{WenbingLin} employed
 the asymptotic expressions of the Hankel function as its
 argument goes to zero. This is difficult to understand because
 what we want to get is the value of the function when its
 argument equals one. For details, please see our discussions in the
 second section.

 In the third section, we will discuss
 effects from the existence of an upper bound on the number of
 e-foldings of inflation on the implementation of the
 initial conditions required of the perturbation mode function.
 If the number of e-foldings of inflation has an upper
 bound, then comoving horizon of the
 universe will have an upper bound as we trace back to
 the origin of the universe. Perturbations at scales greater than
 this bound will keep out of the horizon regardless how
 early we trace back in the history of the
 universe. By assuming that such perturbations
 do not contribute to the angular
 power spectrum of CMBA,  we will find that the observed exceptional
 lowness of the little $l$ multi-poles of CMBA could be reproduced
 numerically. In Linde's $\phi^2$ model, the upper bound on the number of e-foldings of
 inflation takes the value estimated by T. Banks and W. Fischler \cite{Banks1}.

 In our discussions, when specific inflation potentials
 are needed, we will take Linde's chaotic model
 $V(\phi)=m^2\phi^2/2$ as the illustrating examples.
 Except notified explicitly, only density
 perturbations will be discussed in this paper.

 \section{Calculations of the Primordial Power Spectrum of
 Perturbations}

 Taking the simplest single field
 inflation as an example, to get the primordial
 power spectrum of perturbations
 from a given model and make decisive predictions about its
 effects on the large scale structure formation
 in the late time cosmology, we need to solve the
 following coupled Friedmann-Mukhanov equations \cite{StewartLyth, Mukhanov}:
 \beq{}
 &&\ddot{\phi}+3H\dot{\phi}+V^{\prime}(\phi)=0\nonumber\\
 &&H^2=\frac{8\pi G}{3}
 [\frac{1}{2}\dot{\phi}^2+V(\phi)]\nonumber\\
 &&u_k^{\prime\prime}+
 (k^2-\frac{z^{\prime\prime}}{z})u_k=0
 ,\ z=\frac{a\dot{\phi}}{H}\label{Friedmann_Mukhanov001}
 \\
 &&u_k(\eta)|_{k^{-1}/(aH)^{-1}\rightarrow0}\rightarrow
 \frac{1}{\sqrt{2k}}e^{-ik\eta}\label{initial_mukhanov}
 \eeq
 As soon as
 eqs(\ref{Friedmann_Mukhanov001}) are solved,
 spectrum of the primordial
 perturbations can be given as
 \beq{}
 P^{\frac{1}{2}}_{\mathcal{R}}(k)=
 \frac{k^{\frac{3}{2}}}{\sqrt{2}\pi}
 |\frac{u_k(\eta)}{z}|_{aH=k}
 \label{pps_definition}
 \eeq
 Our conventions and formalisms in this paper are
 in agreement with those of \cite{StewartLyth}.

 In the conventional slow roll approximations,
 beginning from,
 \beq{}
 &&\epsilon\equiv\frac{-\dot{H}}{H^2},
 \ \delta\equiv\frac{\ddot{\phi}}{H\dot{\phi}}
 \label{slrparameters}\\
 &&\frac{z^{\prime\prime}}{z}=2a^2H^2(1+\epsilon
 +\frac{3}{2}\delta+\nonumber\\
 &&\hspace{2cm}\frac{1}{2}\delta^2
 +\frac{1}{2}\epsilon\delta
 +\frac{1}{2H}\dot{\epsilon}
 +\frac{1}{2H}\dot{\delta})
 \eeq
 taking $\epsilon$ and $\delta$ as constant,
 eq(\ref{Friedmann_Mukhanov001}) with the initial
 condition eq(\ref{initial_mukhanov}) can be solved analytically,
 \beq{}
 \eta&&\hspace{-3mm}=\int \frac{1}{a}dt
 =\int \frac{1}{a^2H}da\nonumber\\
 &&\hspace{-3mm}=-[\int d\frac{1}{aH}-\int \frac{1}{aH^2}\dot{H}dt]
 \nonumber\\
 &&\hspace{-3mm}=-(aH)^{-1}+\int\epsilon\frac{1}{a}dt
 =-(aH)^{-1}(1-\epsilon)^{-1}
 \label{eta_SRA}\\
 \frac{z^{\prime\prime}}{z}
 &&\hspace{-3mm}
 =\frac{\nu^2-\frac{1}{4}}{\eta^2},\ \text{where}\
 \nu=\frac{1+\epsilon+\delta}{1-\epsilon}+\frac{1}{2}
 \label{zppoz_SRA}\\
 u_k&&\hspace{-3mm}=
 \frac{1}{\sqrt{2k}}
 e^{i(\nu+\frac{1}{2})\frac{\pi}{2}}
 \sqrt{\frac{\pi}{2}}
 (-k\eta)^{\frac{1}{2}}H^{(1)}_{\nu}(-k\eta)\label{stewart28}
 \eeq
 So,
 \beq{}
 P^{\frac{1}{2}}_{\mathcal{R}}(k)
 &&\hspace{-3mm}=\frac{k^{\frac{3}{2}}}{\sqrt{2}\pi}
 |\frac{u_k(\eta)}{z}|_{aH=k}\nonumber\\
 &&\hspace{-3mm}=
 \frac{k^{\frac{3}{2}}}{\sqrt{2}\pi}
 \frac{1}{\sqrt{2k}}\sqrt{\frac{\pi}{2}}H^{(1)}_{\nu}(1)
 [\frac{aH\dot{\phi}}{H^2}]^{-1}_{aH=k}
 \nonumber\\
 &&\hspace{-3mm}=\frac{1}{\sqrt{8\pi}}
 |H^{(1)}_{\nu}(1)|\frac{H^2}{\dot{\phi}}|_{aH=k}
 \label{pps_SRA}
 \eeq
 In eqs(\ref{stewart28}) and (\ref{pps_SRA}), $H_{\nu}^{(1)}(x)$
 is the third class of Bessel function or Hankel function of order
 $\nu$.
 $H_{\nu}(1)$ weakly depends on $k$ through $\nu$, $|H_{1.5}(1)|\approx1.13$.
 We note here that our expressions for
 $P^{\frac{1}{2}}_{\mathcal{R}}(k)$, is
 a little different from that of the reference
 \cite{StewartLyth, WenbingLin} in the numerical factor.
 The authors there used the asymptotical
 expression of eq(\ref{stewart28}) as $k\eta\rightarrow0$
 in calculating their primordial power spectrums,
 which is unnecessary and difficult to understand because
 at the point of horizon exiting, $aH=k,\ k\eta\sim1\nrightarrow0$.
 The same question occurs to the expressions of the
 gravitational wave perturbation power spectrum of reference
 \cite{StewartLyth, WenbingLin}.

 In practice, except in the case of power law inflation,
 in almost all the inflation models, the slow roll parameters
 (\ref{slrparameters}) vary with time. So, eqs(\ref{eta_SRA})-(\ref{pps_SRA})
 can only be taken as the first order approximations. To get
 more precise results, one convenient way is numerical
 calculation. In numerical calculations, eqs(\ref{Friedmann_Mukhanov001})
 should be written in the following first order differential
 equations,
 \beq{}
 \frac{d\phi}{da}&&\hspace{-3mm}=(aH)^{-1}f\nonumber\\
 \frac{df}{da}&&\hspace{-3mm}=-(aH)^{-1}(-3Hf-V^{\prime}(\phi))
 \nonumber\\
 \frac{du_k}{da}&&\hspace{-3mm}=(a^2H)^{-1}w_k\nonumber\\
 \frac{dw_k}{da}&&\hspace{-3mm}=-(a^2H)^{-1}(k^2-\frac{z^{\prime\prime}}{z})u_k
 \nonumber\\
 H^2&&\hspace{-3mm}=\frac{1}{3M_{pl}^2}(\frac{1}{2}f^2+V(\phi))
 \nonumber\\
 \frac{z^{\prime\prime}}{z}&&\hspace{-3mm}=
 a^2[2H^2-V^{\prime\prime}(\phi)-\frac{7}{2M_{pl}^2}f^2
 \nonumber\\
 &&\hspace{10mm}
 -\frac{2}{M^2_{pl}}\frac{fV^{\prime}}{H}+\frac{1}{2M^4_{pl}}\frac{f^4}{H^2}]
 \label{firstorder_fme}
 \eeq
 In the above equations, we have taken scale factor $a$
 instead of time $t$ or conformal time $\tau$ as the
 independent variable. Reference \cite{Salopek} and
 \cite{Adams} use $\text{ln}a$ as the independent variable
 of numerical integration. In principle, using $a$
 or $\text{ln}a$ is equivalent. But in practice, using $a$
 as the independent variable makes
 the stepsize control of the integration routines
 more convenient, so is more time saving.
 This is the first problem we think worthy of notifying in this
 note. The second problem is,
 for little($<10^{-2}Mpc^{-1}$) $k$s,
 we can integrate eqs(\ref{firstorder_fme})
 both by the Fourth Order Runge-Kuta self-adaptive stepsize
 control integration method and the Blosch-Stoer method \cite{Press}. But for
 large $k$s, Runge-Kuta method becomes more and more time-costing
 so that it cannot adapt the variation of our integration goal
 variables at all when $k\approx1Mpc^{-1}$. So, in our final CMBA
 multi-poles' calculation, we use the Blosch-Stoer method to
 calculate the primordial power spectrum of perturbations.

\begin{figure}
 \begin{minipage}[c]{0.8\textwidth}
 \hspace{-5cm}\includegraphics[width=8cm,height=8cm]{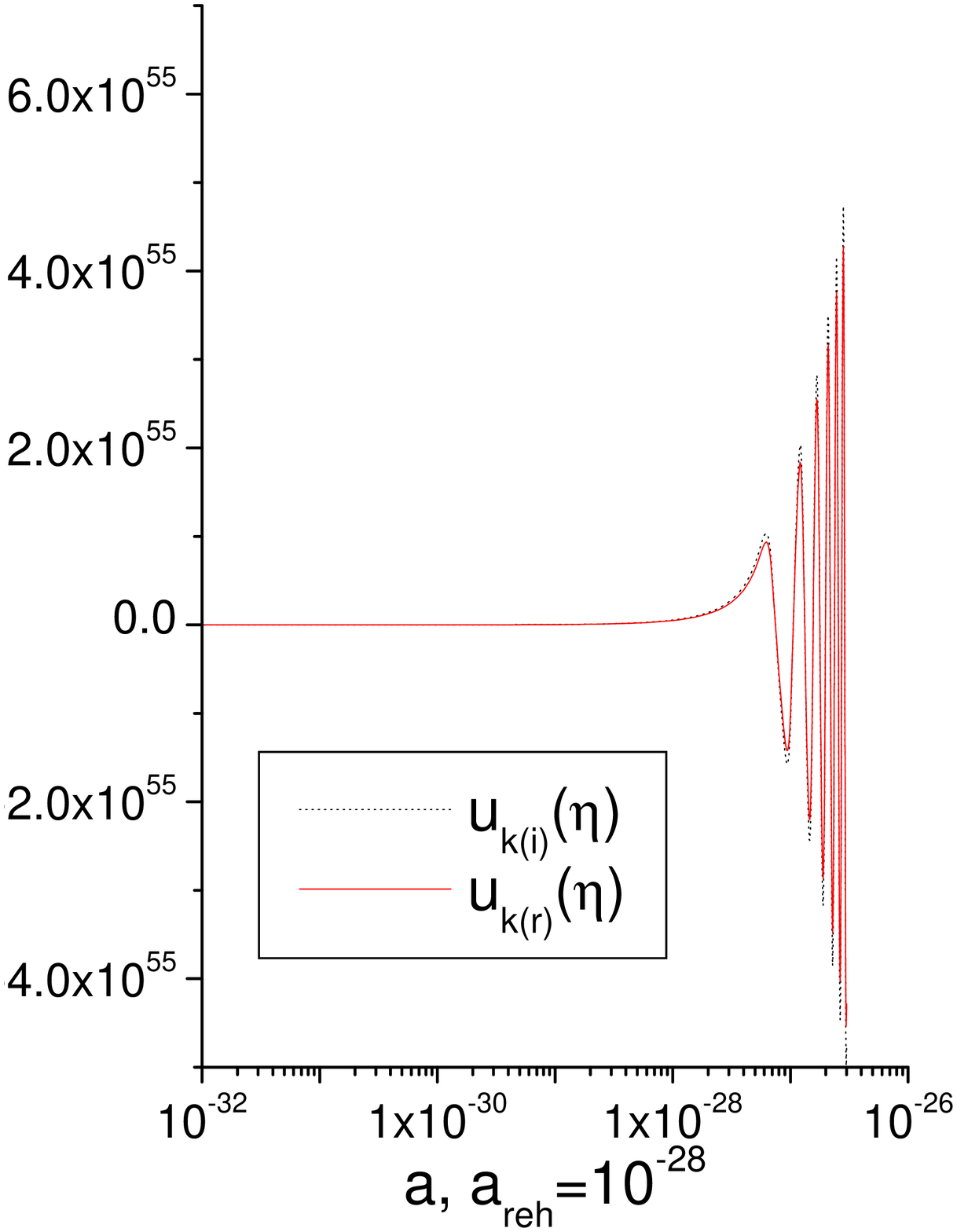}
 \end{minipage}%
 \begin{minipage}[c]{0.1\textwidth}
 \vspace{-27mm}\hspace{-22cm}\includegraphics[width=40mm,height=40mm]{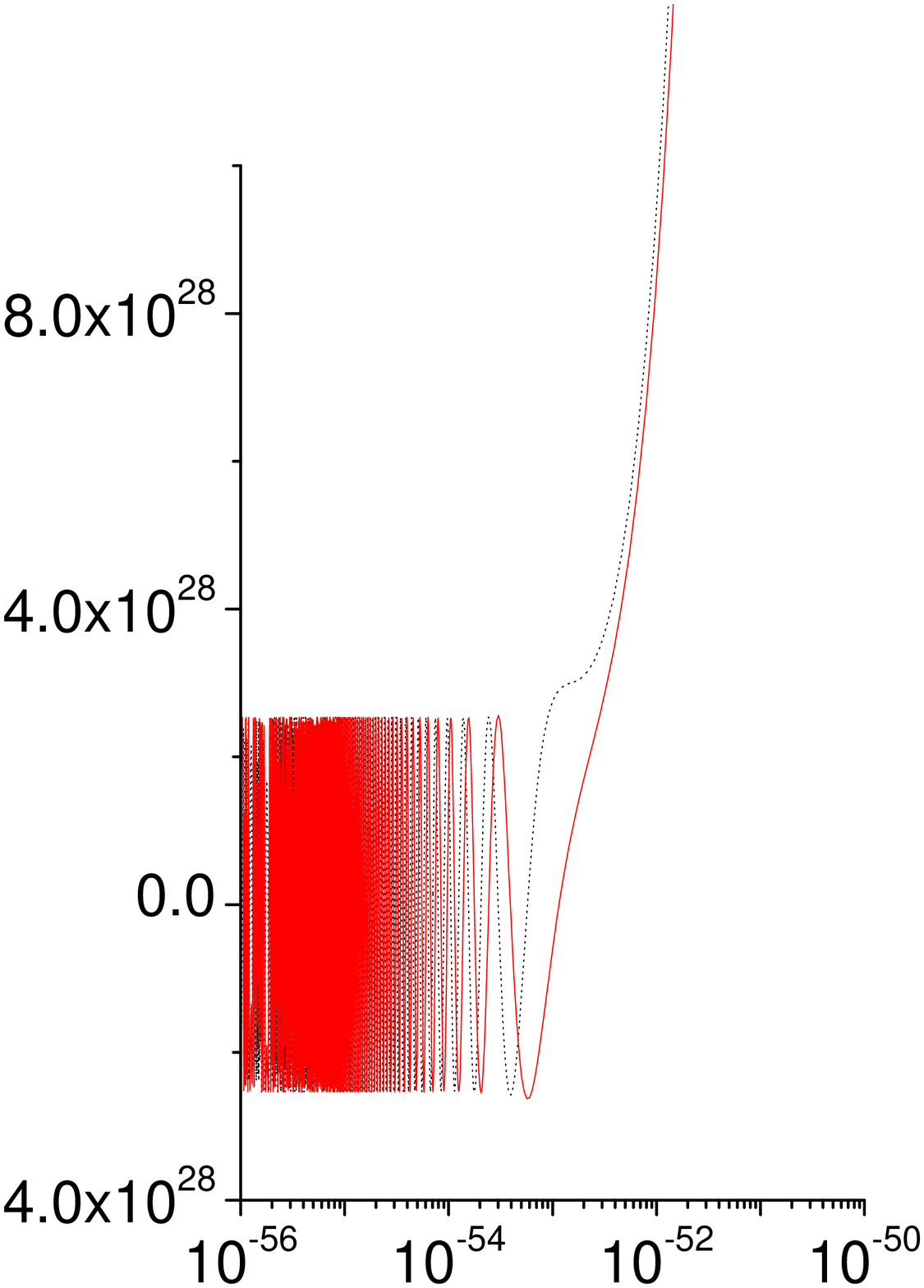}
 \end{minipage}
 \caption{
 Horizon axis, scale factor.
 Vertical axis, mode function $u_{k(i,r)}$,
 imaginary and real part. $k=0.3Mpc^{-1}$.
 $u_k$ oscillates not only at the beginning of inflation but also
 at the end of it. The inflation
 driving potential is $V(\phi)=m^2\phi^2/2$.}
 \label{uk03}
 \end{figure}

\begin{figure}
 \begin{minipage}[c]{0.8\textwidth}
 \hspace{-5cm}\includegraphics[width=8cm,height=8cm]{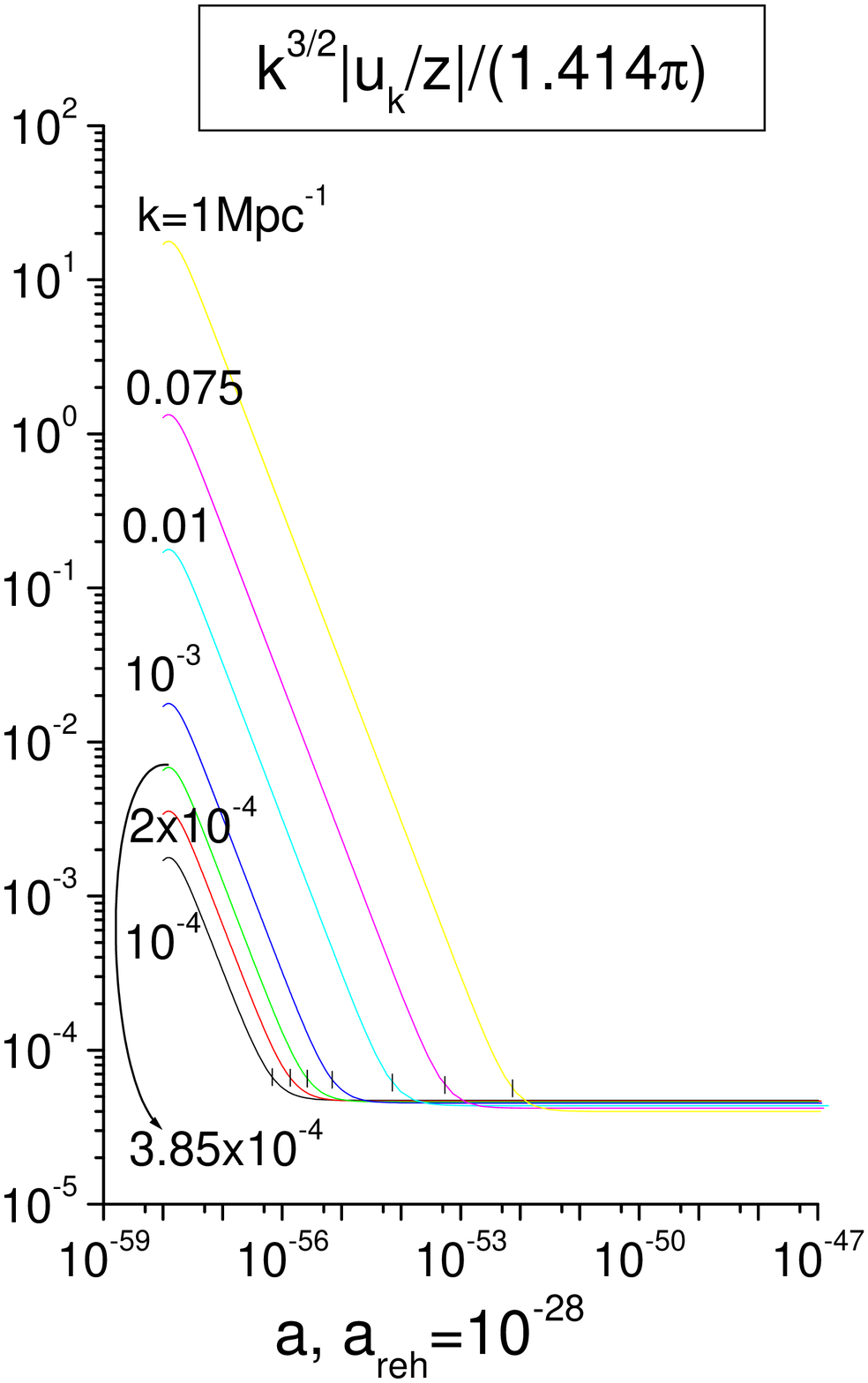}
 \end{minipage}%
 \begin{minipage}[c]{0.1\textwidth}
 \vspace{-2cm}\hspace{-18cm}\includegraphics[width=40mm,height=40mm]{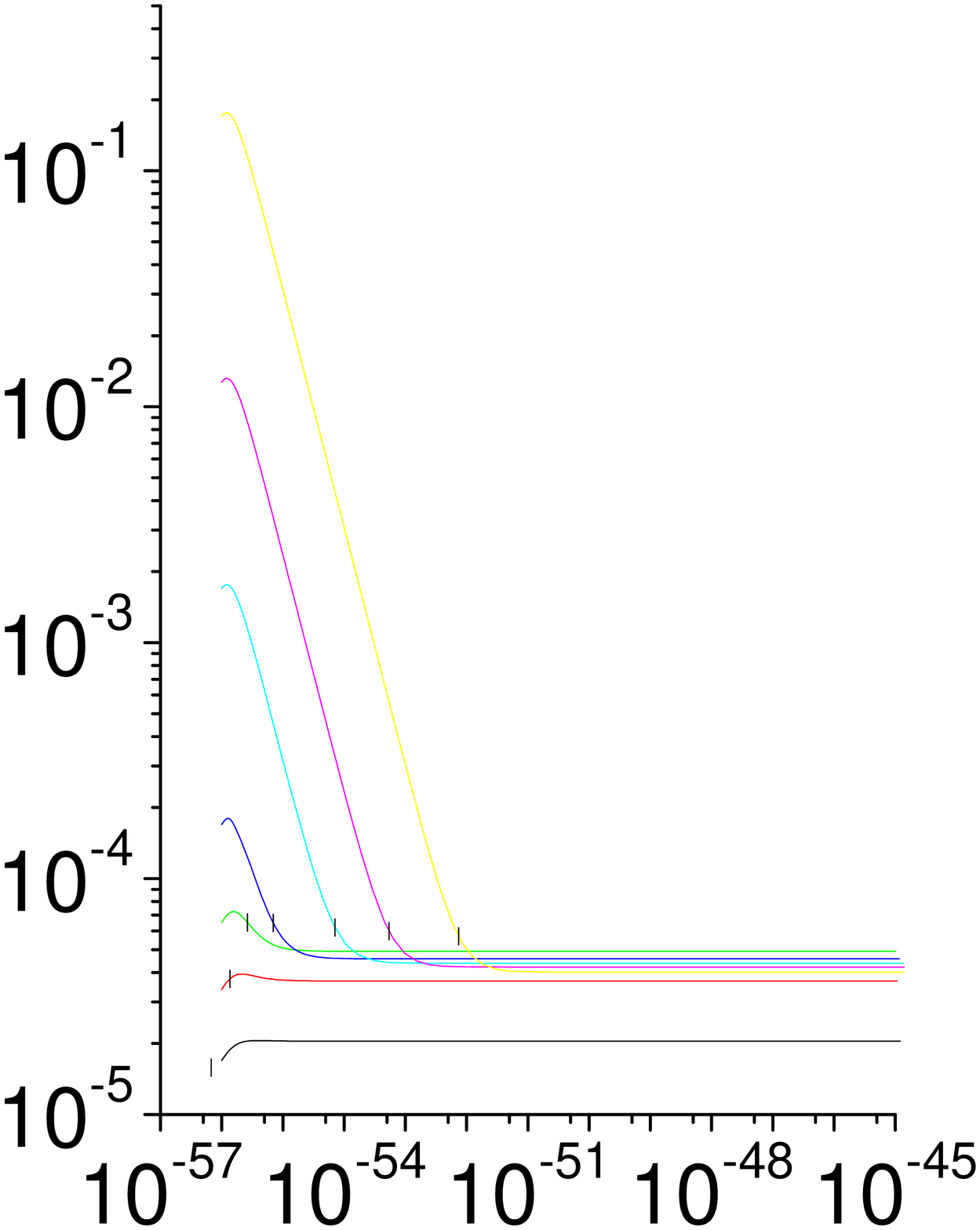}
 \end{minipage}
 \caption{
 Horizon axis: scale factor, normalized by setting $a_{reh}=10^{-28}$.
 Vertical axis: quantity $\frac{k^{3/2}}{\sqrt{2}\pi}|\frac{u_k}{z}|$.
 Different curves corresponds to different $k$.
 Short vertical lines in the figure cross the curves
 at $aH=k$ point. Connecting centers of
 them yields the profile of the primordial power spectrum, see
 also the curves $c$ and $d$ of FIG. \ref{pps}.
 Little figure: obtained by integrating eqs(\ref{firstorder_fme})
 from $a=10^{-56}$ to $a=10^{-28}$, taking eqs(\ref{initial_mukhanov2})
 as the initial conditions.
 Large figure: the same as the little one, but
 the scale factor varies from $10^{-58}$
 to $10^{-28}$ and the initial values of $\phi$ is
 tuned correspondingly so that $10^{30}$ times of inflation
 can be obtained.
 The driving potential is $V(\phi)=m^2\phi^2/2$.}
 \label{k32uoz}
 \end{figure}

 For a given $k$, we give our numerical results of
 the mode function $u_k$ in FIG. \ref{uk03}. From the figure we
 see that the mode function oscillates not only at the beginning
 of inflation, but also at the end of it. It oscillates at
 the end of inflation because $u_k(\eta)\propto\frac{a\dot{\phi}}{H}$
 while $\dot{\phi}$ has oscillation behaviors during this time.
 However the quantity
 $\frac{k^{3/2}}{2\pi^2}|\frac{u_k(\eta)}{z}|$ keeps stable
 after the first period decreasing during inflation,
 see FIG. \ref{uk03} and \ref{k32uoz}. We
 do not know whether the oscillation behavior of $u_k(\eta)$
 at the end of inflation will have any observation implications,
 we note it here only because it was ignored by peoples for long
 time in researches.

 In numeric calculations, we take the integration
 starting point as the conformal time zero point. Using
 the oscillating properties (time translation invariant,
 up to a phase \cite{Salopek, Adams}) of $u_k$ at the beginning of
 inflation,
 the initial conditions can be written as
 \beq{}
 u_{k(ini)}(\eta)=\frac{1}{\sqrt{2k}}e^{-ik\eta}|_{\eta=0}
 \nonumber\\
 w_{k(ini)}(\eta)=\sqrt{\frac{k}{2}}e^{-ik\eta}|_{\eta=0}
 \label{initial_mukhanov2}
 \eeq
 To solve the horizon, flatness and other
 pre-inflation problems,
 we require the universe expand $N_e=65$ e-foldings during
 inflation. The initial values of the scalar field has been
 tuned so that 65 e-foldings of inflations can be obtained.
 Assuming the reheating temperature is about
 $10^{15}Gev$ and the universe expand adiabatically
 in the standard big bang process, then at the initial
 point of integration, the scale factor is
 \beq{}
 a_{ini}=[e^{N_e}\frac{T_{reh}}{T_{tod}}]^{-1}=10^{-56},
 \eeq
 provided that the scale factor of today is set to $1$.
 However, if we let $k$ varies from $[10^4Mpc]^{-1}$ to $1Mpc^{-1}$
 and integrate eqs(\ref{firstorder_fme})
 from $a=10^{-56}$ to $a=10^{-28}$
 to get the full power spectrum, what we find will be a large scale
 suppressed result, see FIG. \ref{k32uoz}, \ref{pps} and the notes there.

 The observational result of
 \cite{WMAP1}, if verified, favors a large scale suppressed power spectrum,
 but our suppressions here
 is due to an inconsistent initial conditions imposing.
 Because for very
 little $k$s, at the integration starting point $a_{ini}=10^{-56}$,
 $k^{-1}/(aH)^{-1}_{ini}\nrightarrow0$. So eqs(\ref{initial_mukhanov2})
 cannot be taken as the correct initial conditions. For example,
 if we want to integrate eqs(\ref{Friedmann_Mukhanov001}) in
 the simplest chaotic inflation model,
 $V(\phi)=m^2\phi^2/2$, $m=5.61\times10^{-6}M_{pl}$ for $k=H_0=(4225Mpc)^{-1}$.
 To assure driving the universe expanding $10^{28}$ times during
 inflation, $\phi_{ini}=16.2M_{pl}$ \cite{BunnLiddleWhite}.
 Then, at the integration starting point,
 \beq{}
 (aH)^{-1}_{ini}
 &&\hspace{-3mm}=[a_{ini}\sqrt{\frac{1}{3M^2_{pl}}V(\phi_{ini})}]^{-1}
 \nonumber\\
 &&\hspace{-3mm}=[10^{-56}\times37.1\times10^{-6}M_{pl}]^{-1}
 \sim7082Mpc\nonumber\\
 \label{horizon_criteria}
 \eeq
 Obviously, $k^{-1}/(aH)^{-1}_{ini}\sim0.60\nrightarrow0$, i.e.
 this $k$ mode of fluctuation is not well inside the horizon of
 that time.
 So, to use eqs(\ref{initial_mukhanov2}) as the initial conditions,
 we have to integrate eqs(\ref{Friedmann_Mukhanov001}) or
 equivalently eqs(\ref{firstorder_fme})
 beginning from a more little scale factor,
 say $a_{ini}=10^{-58}$ for example, so that those modes of fluctuations
 locate well inside the horizon of the integration starting point. Of course,
 the initial values of $\phi$ should be tuned correspondingly so that
 $10^{30}$ times of inflation can be obtained.
 Although works on the numerical calculations of the primordial power
 spectrum existed in several literatures \cite{Salopek, Adams, WenbingLin},
 none of them pointed out this problem explicitly,
 so we think it is worth noting here.

 If we only let the universe expand
 $10^{26}\approx e^{60}$ \cite{BunnLiddleWhite} times during
 inflation, the problem discussed above would be more
 serious.

 \section{The Upper Bound on the Number of E-foldings of Inflation}

 However, if the number of e-foldings of inflation has an
 upper bound $N_{u.b.}$, then the comoving horizon of the universe will
 have an upper bound also. So as we trace back beyond this upper
 bound, the comoving horizon will decrease. Perturbations at
 scales larger than this upper bound will keep out of
 the horizon regardless how early we trace back to the
 origin of the universe. For such perturbations,
 the initial conditions eq(\ref{initial_mukhanov})
 cannot be implemented physically.

 Then, what is the upper bound? Is it
 large enough so that, at least for perturbations
 at scales of today's Hubble horizon, the initial conditions
 eqs(\ref{initial_mukhanov}) can be implemented numerically?
 In string theories, by T. Banks and W. Fischler's argument \cite{Banks1},
 if the present acceleration of the universe is due to an
 asymptotically de Sitter universe with small cosmological
 constant, and the principle of Cosmological Complementary is
 valid, then the number of e-foldings of inflation is bounded.
 They estimated that $N_{u.b.}=65$. So the value already equals
 our assumed number of e-foldings of inflation required to
 solve the flatness, horizon and other pre-inflation
 problems in eqs(\ref{horizon_criteria}). Hence, if T. Banks and
 W. Fischler's estimation
 is the case, then at least for perturbations at scales of today's
 Hubble horizon, it will be impossible to implement the initial
 conditions (\ref{initial_mukhanov}) physically. Hence the power
 spectrum of such perturbations could not be nearly scale free
 any more. It could be suppressed strongly comparing with that
 of the little scale perturbations.

 The second question is, if the number of e-foldings of inflation,
 consequently, the comoving horizon as we trace back to the origin
 of the universe has an upper bound, then,
 do perturbations at scales larger than this bound
 contribute to our observed CMBA?
 Recalling \cite{WMAP2} that, in calculating the multi-poles of CMBA, we
 need to do the following integration:
 \beq{}
 C_{l}=\int_0^{\infty}\frac{dk}{k}P_{\mathcal{R}}(k)[g_{l}^{\mathcal{R}}(k)]
 \label{integration_cl}
 \eeq
 where $P_{\mathcal{R}}(k)$ is the primordial power spectrum of
 perturbations produced during inflation and
 $g_l^{\mathcal{R}}(k)$ is the radiation transfer function.
 Mathematically, the integration should be made from $0$ to
 $\infty$. But physically, if some mechanism exists so that
 perturbations at scales greater than
 a certain cutoff wavelength $k_c^{-1}$
 do not contribute to CMBA, then the lower bound of the
 integration (\ref{integration_cl}) will be changed unavoidably.

 \begin{figure}
 \includegraphics[width=7cm,height=8cm]{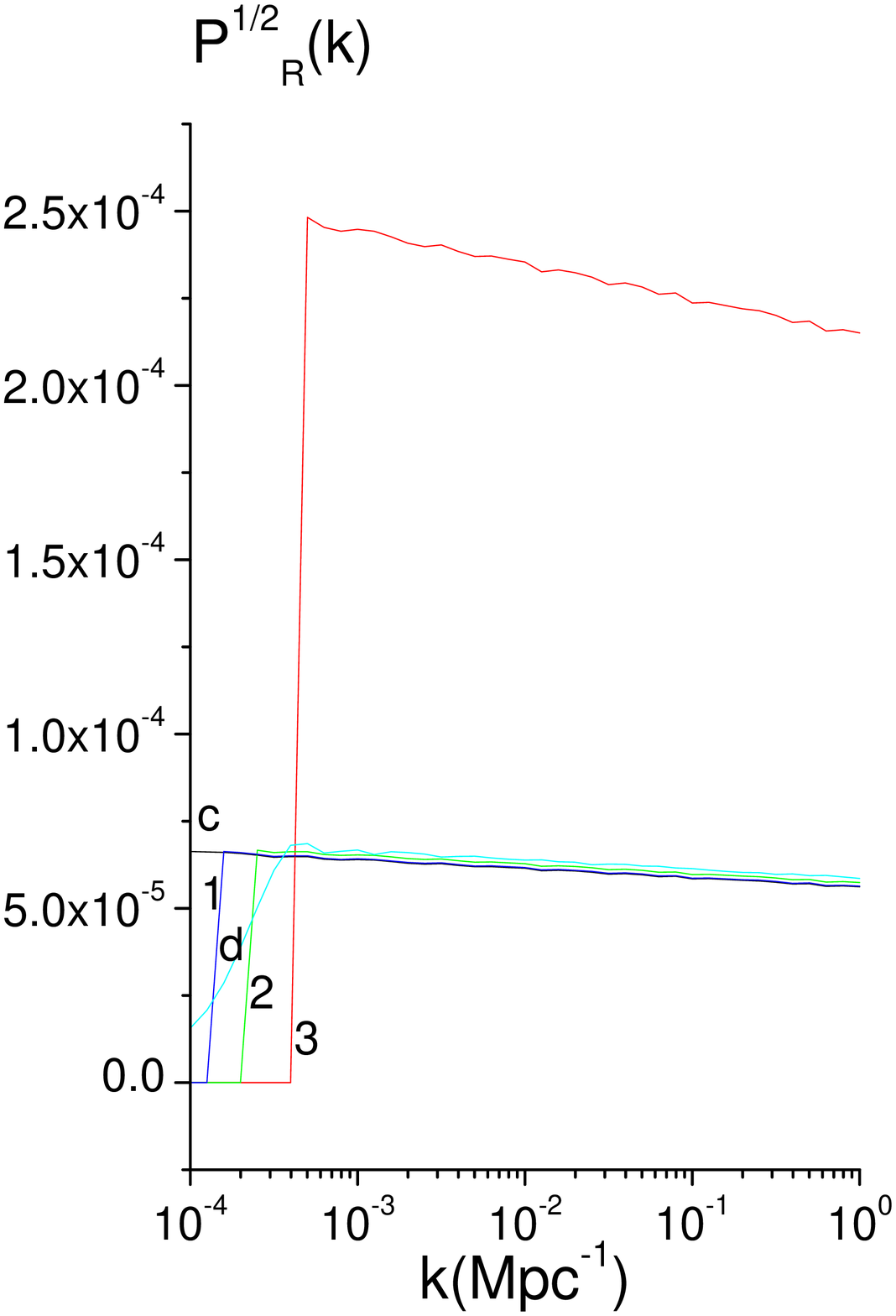}
 \caption{Primordial power spectrum of density perturbations
 in Linde's $\phi^2$ model.
 c: no suppression at all.
 d: large scale power suppressed by the inconsistent initial conditions.
 \newline
 1: by assuming that $P_\mathcal{R}(k)=0$ for all $k<[7082Mpc]^{-1}$.
 \newline
 2: by assuming that $P_\mathcal{R}(k)=0$ for all $k<[4225Mpc]^{-1}$.
 \newline
 3: by assuming that $P_\mathcal{R}(k)=0$ for all $k<[2113Mpc]^{-1}$.
 }
 \label{pps}
 \end{figure}
 \begin{figure}
 \includegraphics[width=7cm,height=8cm]{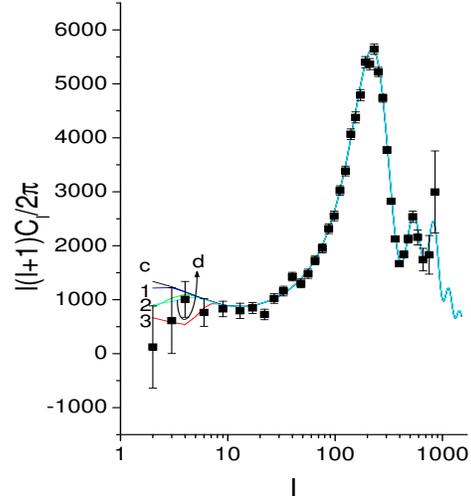}
 \caption{Theoretical curves is computed by CMBfast.
 Except the primordial power spectrum, see
 FIG. \ref{pps} and the notes there, other cosmic parameters
 are from WMAP's best fitting $\Lambda CDM$ model.
 Scattered points are the observational results of WMAP2003  \cite{WMAP1}.
 }
 \label{Comparisons}
 \end{figure}

 Obviously, if the
 number of e-foldings of inflation has an upper bound, then at the
 beginning of inflation, comoving horizon of the universe
 will reach its upper bound $[a_{i.b.}H_{i.b.}]^{-1}$ as we
 trace back to the origin of the
 universe. Perturbations beyond such a bound will keep out of the horizon
 regardless how early we trace back. We propose that such perturbations
 do not contribute to our observed CMBA.
 Our motivation is: from the beginning of inflation, these perturbations lie outside
 the horizon. They have no chances to experience a transition from
 quantum ones to classical ones before CMBA formed.

 Under this assumption, we write the primordial power spectrum of
 perturbations produced during inflation as
 \beq{}
 P_\mathcal{R}(k)&&\hspace{-3mm}=0,\ k<k_c
 \nonumber\\
 &&\hspace{-3mm}=A\frac{k^3}{2\pi^2}|\frac{u_k}{z}|^2_{aH=k},
 \ k\geq k_c
 \label{cutoff_prk}
 \eeq
 where $A$ is a constant whose
 value can be determined by requiring that the
 corresponding primordial power
 spectrum gives approximately the same first peak
 CMBA multi-poles as observations,
 see FIG. \ref{pps} and \ref{Comparisons},
 while $k_c$ is determined by the following relation:
 \beq{}
 k_c^{-1}=[a_{i.b.}H_{i.b.}]^{-1}\label{kc_definition}
 \eeq
 where the subscript $i.b.$ means that the relevant quantity take value when inflation
 begins.
 \beq{}
 a_{i.b.}&&\hspace{-3mm}=[e^{N_{u.b.}}\frac{T_{reh}}{T_{tod}}]^{-1}\nonumber\\
 H^2_{i.b.}&&\hspace{-3mm}\approx\frac{\Lambda_{infl}}{3M_{pl}^2}\nonumber\\
 k_c^{-1}&&\hspace{-3mm}\approx e^{N_{u.b.}}\frac{T_{reh}}{T_{tod}}
 \frac{\sqrt{3}M_{pl}}{\sqrt{\Lambda_{infl}}}
 \label{horizon_upperbound}
 \eeq
 In the above equations, the scale factor of today is set to one,
 $N_{u.b.}$ denotes the upper bound on
 the number of e-foldings of inflation, $T_{reh}$ and $T_{tod}$
 are the temperature of the universe at the reheating point and today
 respectively, while $\Lambda_{infl}$ is the characteristic energy
 scale of a given inflation model. We supposed that the universe enter
 the standard adiabatic expansion immediately after inflation.

 Take T. Banks and W. Fischler's estimation as an
 input and eqs(\ref{horizon_criteria}) as
 a guide post, where by the logic of eqs(\ref{horizon_upperbound})
 $N_{u.b.}=65$, $T_{reh}\approx10^{15}Gev$,
 $T_{tod}\approx10^{-4}ev$, $\Lambda_{infl}=[m^2\phi^2/2]^{1/4}=8.01\times10^{-3}M_{pl}$,
 we numerically calculated the CMBA multi-poles for three values of $k_c$
 \beq{}
 k_{c1,2,3}=[7082Mpc]^{-1},[4225Mpc]^{-1},[2113Mpc]^{-1}
 \label{three_kc}
 \eeq
 where $7082Mpc$ could be considered as the largest sub-horizon perturbation scale
 under the assumption that $N_{u.b.}=65$; $4225Mpc$ is the perturbation scale
 corresponding to our today's hubble horizon; while $2113Mpc$, half of
 that.
 Our results are displayed in
 FIG. \ref{pps} and \ref{Comparisons}, curves 1, 2, 3.
 From the figure we
 see that the little $l$ multi-poles of CMBA
 is very sensitive to the cutoff of the primordial power spectrum.
 If $k_c$ takes a certain value between $k_{c2}$ and $k_{c3}$,
 then theoretical curves will
 tally with the observational results to a better degree.

 So, if the number of e-foldings of inflation has an upper bound,
 and by our proposition, perturbations outside the horizon at the
 inflation beginning point do not affect the large scale structure
 formation at late times, then the exceptional lowness of the little $l$
 multi-poles of CMBA can be reproduced numerically. In Linde's
 $\phi^2$ inflation model, which has characteristic energy
 $\Lambda\approx10^{-2}M_{pl}$, the upper bound on the number of
 e-foldings of inflation is approximately $65$. In other inflation models,
 if the observational result are to be reproduced,
 by eq(\ref{horizon_upperbound}), more lower characteristic inflation energy
 will mean a more little upper bound on the number of e-foldings,
 and vice versa.

 When this paper is posted in the network, we are noted that
 in reference \cite{Stochastic}, the authors
 studied the stochastic inflation mechanism and got similar
 results like ours. In some sense, the mechanisms there
 can be assimilated to another statement of a limited number of
 e-foldings of inflation.

\vspace{-5mm}\section{Summary}

In this paper, we first present four notes on calculations of the
primordial power spectrum of perturbations produced during
inflation. 1st, to get the nearly scale free power spectrum of
perturbation, it is unnecessary to use the asymptotic behavior of
the Hankel function as its argument goes to 0. 2nd, scale factor
$a$ is a better independent variable than $lna$ in the numerical
integrations of the Friedmann-Mukhanov equations. 3rd, the
perturbation mode function $u_k$ oscillates both at the initial
and the end of inflation times. 4th, numerically, to implement
the initial conditions of the perturbation equations
consistently, we must let the universe experience more number of
e-foldings inflation than that required of it to solve the
horizon, flatness and other pre-inflation problems.

We then explore effects from the existence of an upper bound on
the number of e-foldings of inflation on the implementation of
the initial conditions required of the perturbation mode
function. When such an upper bound exists, the comoving horizon
of the universe will have an upper bound also as we trace back to
the origin of the universe. So perturbations at scales beyond this
bound will keep outside the horizon regardless how early we trace
back in the history of the universe. After proposing that such
perturbations do not affect the formation of CMBA, we find the
observed exceptional lowness of the little $l$ multi-poles of
CMBA could be reproduced numerically. In Linde's $\phi^2$ model,
the upper bound on the number of e-foldings of inflation is
determined to be 65 approximately.

{\bf{Acknowledgments:}} We would like to thank professor M. Li for
discussions.

%%%%%%%%%%%%%%%%%%%%%%%%%%%%%%%%%%%%%%%%%%%%%%%%%%%%%%%%%%%%%%%%%%%%%%%

\end{document}